\documentclass{mekit19}
\usepackage{xcolor}
\usepackage{graphicx}
\usepackage{amsmath}
\usepackage{amsfonts}
\usepackage{amssymb}
\usepackage{lmodern}
\usepackage{cleveref}
\usepackage{subcaption}

\newcommand{\pdts}[1]{\partial_t #1}
\newcommand{\pdt}[1]{\partial_t\left(#1\right)}

\newcommand{\pdis}[1]{\partial_i #1}

\newcommand{\pdjs}[1]{\partial_j #1}
\newcommand{\pdj}[1]{\partial_j\left(#1\right)}
\newcommand{\pdks}[1]{\partial_k #1}
\newcommand{\pdk}[1]{\partial_k\left(#1\right)}

\newcommand{\vavg}[1]{\overline{#1}}
\newcommand{\phavg}[1]{\langle #1 \rangle}
\newcommand{\favg}[1]{\tilde{#1}}
\newcommand{\wfavg}[1]{\widetilde{#1}}

\newcommand{\phrho}{\phavg{\rho}}

\newcommand{\php}{\phavg{p}}

\newcommand{\fur}{\favg{u}_r}

\newcommand{\pdr}[1]{\partial_r\left(#1\right)}

\newcommand{\Rrr}{\favg{R}_{r r}}
\newcommand{\Rtt}{\favg{R}_{\theta \theta}}

\title{Numerical investigation of shock wave particle cloud interaction in cylindrical geometries}

\author{Andreas N. Osnes$^{1}$, Magnus Vartdal$^{2}$, Marianne G. Omang $^{3,4}$, and Bj\o rn A. P. Reif$^{1}$}

\heading{Andreas N. Osnes, Magnus Vartdal and Bj\o rn A. P. Reif}

\address{
$^{1}$ Department of Technology Systems, University of Oslo (UiO) \\
PO Box 70 Kjeller, 2027 Kjeller\\
e-mail (A. N. Osnes, corresponding author): a.n.osnes@its.uio.no \\
e-mail (B. A. P. Reif): b.a.p.reif@its.uio.no
\and
$^{2}$Norwegian Defence Research Establishment (FFI)\\
PO Box 25 Kjeller, 2027 Kjeller\\
e-mail: magnus.vartdal@ffi.no 
\and
$^{3}$Norwegian Estates Research Agency \\
PO Box 405 Sentrum, 0103 Oslo
\and
$^{4}$Institute of Theoretical Astrophysics, University of Oslo (UiO)\\
PO Box 1029 Blindern, 0315 Oslo\\
e-mail: m.g.omang@astro.uio.no
\and
\textbf{Preprint submitted for consideration for publication in the proceedings of MekIT'19 - 10th National Conference on Computational Mechanics, 3rd-4th June 2019, Trondheim,  Norway}
}
\abstract{This study investigates the interaction of a shock wave with a fixed layer of particles in cylindrical geometries using particle-resolved large eddy simulations. The curvature radius of the particle layer is varied and the resulting flow variations are analyzed. The mean flow field depends strongly on the curvature radius, but this is not the case for flow fluctuations or particle drag coefficients. The results indicate that particle scale flow phenomena are insensitive to geometric expansion within the range investigated here. This is an encouraging result from a modelling perspective, since it means that results and observations of particle scale flow phenomena obtained in planar configurations can likely be extrapolated to diverging geometries.}
\begin{document}
\section{INTRODUCTION}
The interaction of shock waves with particle clouds plays an important role in a number of natural phenomena, industrial applications, and safety measures such as volcanic eruptions \cite{bower1996}, shock wave mitigation using porous barriers \cite{suzuki2000,chaudhuri2013}, and ejection of stellar dust from supernovae \cite{silvia2012}. The primary motivation for the present work is the role of shock wave particle cloud interaction in heterogeneous explosives \cite{zhang2006} and explosive dissemination of powders and liquids \cite{zhang2001,milne2010,rodriguez2017}. The latter applications typically include significant geometric expansion effects. The effect of this expansion on the interaction process is the topic of this work.    

Dispersal of cylindrical particle shells by shock waves has previously been studied experimentally using both explosives \cite{milne2010,zhang2001,frost2012} and shock tubes \cite{rodriguez2014,rodriguez2017}. In both cases, the particle layers have initially been very dense, with volume fractions approaching the random packing limit. The initial dispersion of these shells are therefore subject to strong particle collision effects. As the powders are accelerated outward, the particle volume fraction quickly decreases as a result of geometric expansion. Consider for instance a layer of initial thickness $L$ and inner curvature radius $R_0=L$ that initially has a particle volume fraction, $\alpha_\mathrm{p}$, of $0.5$. If this layer is accelerated outwards without changing thickness, it will have $\alpha_\mathrm{p}=0.3$ when the inner radius is $2L$, and $\alpha_\mathrm{p}\approx0.2$ at $R_0=3L$. Thus, despite having a large initial volume fraction, the particle rapidly enter the intermediate volume fraction regime ($\alpha_\mathrm{p}=0.01-0.5$) and remain there for a substantial part of the dispersal process in such scenarios. Furthermore, one of the primary flow features of this dispersal process is the formation of particle jets \cite{frost2012,frost2018}. The preferential concentration of the particles in jets tends to increase the time during which particles remain in the intermediate volume fraction regime. During this time, the flow periodically over-expands. This results in implosions that generate secondary shock waves which propagate outwards through the particle cloud. Interactions between shock waves and particle clouds in the intermediate volume fraction regime is therefore one of the primary flow features in this dispersal process.

From a modelling perspective, the intermediate volume fraction regime is especially challenging because the dynamics are affected by a complex interaction between the flow field and the particle distribution \cite{theofanous2017-2}. Each particle interacts with the incoming shock wave and subsequent flow in a manner that depends on the local particle configuration. The interaction generates reflected shocks, shear layers, and wakes that in turn interact with nearby flow features. These flow perturbations result in large particle drag force variations that alter the configuration of particles. Consequently any modelling effort where the interaction between the particles and the flow is assumed to consist of a sum of interactions with isolated particles is unlikely to succeed. It is therefore necessary to perform detailed investigations of the flow around and forces on the particles, and relate the observations to available model quantities, in order to establish suitable simplified models for this regime. 

Experimental investigation of flow features at the particle scale inside particle clouds is challenging. It is, however, possible to conduct particle-resolved simulations for this purpose, and several recent studies have done just that \cite{regele2014,hosseinzadeh2018,sen2018,mehta2018,theofanous2018,vartdal2018,osnes2019}. Such simulations are computationally expensive, since a large number of particles must be used to obtain meaningful statistics. In addition to yielding physical insight, particle resolved simulations can be used to investigate closures for unresolved terms that appear in simpler dispersed flow models. The unclosed terms are a result of averaging of products of fluctuations. What the fluctuation products represent depends on the averaging type. For shock wave particle cloud interaction it is convenient to apply volume averaging. With this approach, both turbulent fluctuations and laminar flow effects around particles, often referred to as pseudo-turbulent fluctuations, contribute to fluctuation correlations. In addition, new terms appear due to averaging over volumes containing gas and particles, as discussed in e.g. \cite{schwarzkopf2015}. Particle-resolved simulation data can be utilized to examine all of these terms. This is particularly useful for development of Eulerian-Eulerian and Eulerian-Lagrangian dispersed flow models \cite{shallcross2018,theofanous2017-2,mcgrath2016,saurel2017}. In addition, the results of resolved simulations can be used directly as validation data for the simplified models.

In this work, we examine the effect of flow expansion on the passage of shocks through particle clouds in the intermediate volume fraction regime ($\alpha_p=0.1$). Flow expansion causes rapid spatial variation of mean flow fields, and this work explores how this affects the flow through particle clouds. We conduct particle resolved large eddy simulations of a shock wave passing through a cylindrical shell of randomly positioned stationary particles. We vary the radius of curvature of the cylindrical shell and keep the shell thickness constant. For each curvature radius we perform an ensemble of simulations to obtain statistically representative results.   

This paper is organized as follows. In \Cref{sec:goveq} the governing equations and the volume averaged equations used for analysis are presented. \Cref{sec:comp-set-up} describes the computational method and the set-up of the problem. \Cref{sec:grid-conv} contains results from grid and ensemble convergence studies. \Cref{sec:results} presents the simulation results.
We examine wave trajectories, mean flow fields, flow fluctuations and particle forces. We also investigate the relative importance of the terms in the volume averaged momentum equation in different regions.
Finally, concluding remarks are given in \cref{sec:conclusions}.

\section{GOVERNING EQUATIONS}
\label{sec:goveq}
The governing equations for the gas dynamics in this work are the conservation equations of mass, momentum and energy
\begin{equation}
\pdts{\rho}+\pdk{\rho u_k}=0,
\label{eq:mass}
\end{equation}
\begin{equation}								
\pdt{\rho u_i}+\pdk{\rho u_i u_k}=-\pdis{p} + \pdjs{\sigma_{ij}},
\label{eq:momentum}
\end{equation}
\begin{equation}
\pdt{\rho E}+\pdk{\rho E u_k + p u_k}=\pdj{\sigma_{ij}u_i} - \pdk{\lambda\pdks{T}}.
\label{eq:energy}
\end{equation}
Here, $\rho$ is the mass density, $u$ is the velocity, $p$ is the pressure, $\sigma_{ij}=\mu (\pdjs{u_i}+\pdis{u_j}-2\pdks{u_k}\delta_{ij}/3)$ is the viscous stress tensor, $E=\rho e + 0.5\rho u_ku_k$ is the total energy per unit volume, $\lambda$ is the thermal conductivity, $T$ is the temperature,  $\mu$ is the dynamic viscosity and $e$ is the internal energy per unit mass. A calorically perfect ideal gas equation of state with $\gamma=1.4$ is employed. Furthermore, we assume a power law dependence of viscosity on temperature with an exponent of $0.76$ and a constant Prandtl number of $0.7$. 

The above equations are inconvenient for analysis of the simulation results, due to the strong spatial variation within the particle cloud. Instead, the volume averaged equations of motion are used for analysis. These are obtained by averaging \cref{eq:mass,eq:momentum,eq:energy} over a volume, and carefully accounting for the effect of the dispersed phase within that volume. In this study, $\vavg{\cdot}$ will be used to denote volume averaging, $\langle\cdot\rangle$ denotes phase-averaging, and $\favg{\cdot}$ denotes Favre-averaging. The deviations from Favre-averaged values are denoted by $\cdot''$. Phase and volume averaging are related by $\alpha\phavg{\cdot} = \vavg{\cdot}$, where $\alpha$ is the gas phase volume fraction. The problem under consideration is statistically homogeneous in the axial ($z$) and azimuthal ($\theta$) directions. Therefore, the only component of the volume averaged momentum equation that is of interest is the radial ($r$) one. Assuming stationary inert particles, the volume averaged mass and momentum equations in the radial direction are 
\begin{equation} \label{eq:Vmass}
\pdt{\alpha \phrho} + \pdr{\alpha \phrho \fur} = -\frac{\alpha\phrho\fur}{r},
\end{equation}
\begin{equation} \label{eq:Vmom}
\begin{split}
\pdt{\alpha \phrho\fur} + \pdr{\alpha\phrho\fur\fur  + \alpha \php}=-\frac{\alpha\phrho}{r}\fur\fur +\pdr{\alpha\phavg{\sigma_{rr}}}+ \frac{\alpha\phavg{\sigma_{rr}}}{r}&\\
-\pdr{\alpha\phrho\favg{R}_{rr}}
- \frac{\alpha\phrho}{r}\favg{R}_{rr} +\frac{1}{V}\int_Sp n_r dS - \frac{1}{V}\int_S \sigma_{rk}n_kdS.&
\end{split}
\end{equation}
Here, the boundary between the gas and the particles is denoted by $S$, $V$ is the averaging volume and $n_k$ is the particle surface normal. The integrals represent the forces acting on the particle surfaces. $\favg{R}_{rr}=\wfavg{u_r''u_r''}$ is the radial component of the average stress due to velocity fluctuations. 
The full tensor, $\favg{R}_{ij}$, is the single-point, density weighted (Favre averaged), velocity fluctuation correlations, i.e.
\begin{equation}
\favg{R}_{ij}=\frac{\phavg{\rho u_i''u_j''}}{\phavg{\rho}}.
\end{equation} 
$\favg{R}_{ij}$ contains both the classical turbulent stresses and the pseudo turbulent stresses mentioned in the introduction. Its role is analogous to that of the classical Reynolds stress in the RANS equations and we thus refer to this term as Reynolds stress in the rest of the paper.

\section{COMPUTATIONAL METHOD AND SET-UP}
\label{sec:comp-set-up}

\subsection{Computational method}
The governing equations are solved numerically using the compressible flow solver "CharLES" from Cascade Technologies. It employs an entropy stable scheme on a Voronoi-mesh with third order Runge-Kutta time stepping \cite{bres2018}. For further discussions of entropy stable schemes, consult e.g. \cite{tadmor2003,chandrashekar2013}.

\subsection{Problem set-up}
\label{sec:problemsetup}
\begin{figure}
	\centering
	\includegraphics{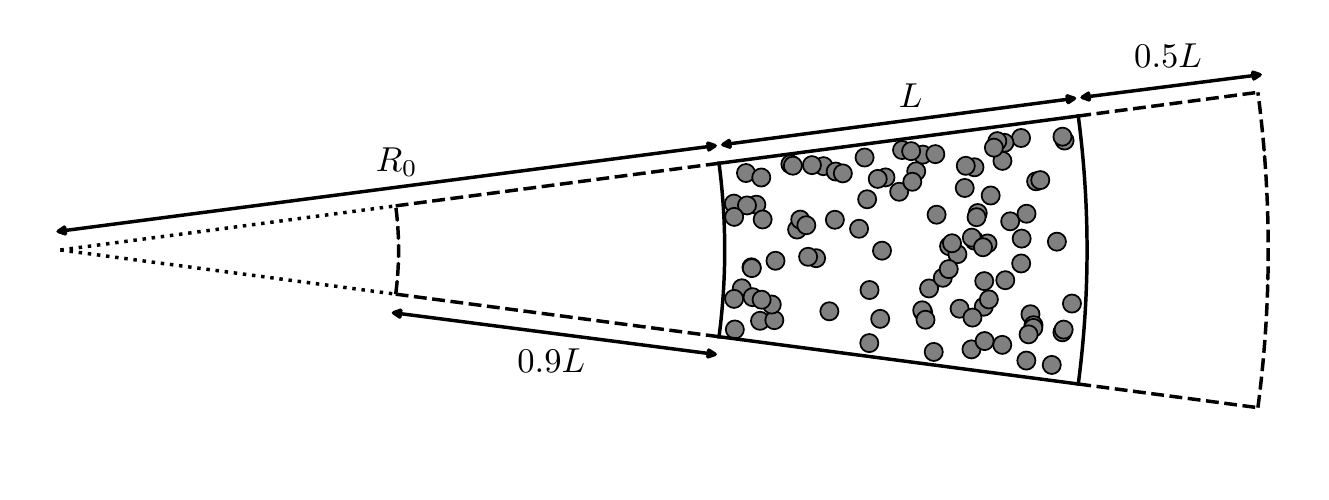}
	\caption{Sketch of the computational domain. The dashed lines indicate the domain being simulated. $R_0$ is the radius of curvature and $L=1.2\sqrt[3]{4}\ \mathrm{mm}$ is the particle layer thickness. The particle cloud is located between $R_0$ and $R_0+L$. The particle diameter is $D_\mathrm{p}=4^{-1/3}\times10^{-1}\ \mathrm{mm}$.}
	\label{fig:sketch}
\end{figure}

This work considers the effect of geometric expansion on the passage of shocks through particle clouds. To this end, we conduct simulations in a cylindrical geometry, where the particle cloud is a cylindrical shell. We consider spherical particles with diameter $D_\mathrm{p}=4^{-1/3} \times10^{-1}\ \mathrm{mm}$. The shell-thickness is denoted $L$, and is kept constant at $1.2\sqrt[3]{4}\ \mathrm{mm}\simeq 30.2D_\mathrm{p}$. We consider three different radii of curvature defined such that the inner particle shell radius, $R_0$, takes the values $L$, $2L$ and $\infty$. \Cref{fig:sketch} shows a sketch of the computational domain.  The arc-length of the inner shell edge is kept constant at $8\sqrt[3]{4}D_\mathrm{p}$ by considering cylindrical sectors with different angles. In the axial direction, a constant domain size of $8\sqrt[3]{4}D_\mathrm{p}$ is used. For each curvature radius, the inner boundary is located $0.9L$ upstream of the particle shell edge.

The particles occupy a volume fraction $\alpha_p=0.1$ within the layer, and their positions are drawn randomly with the condition that none of the particles are closer than $0.05D_\mathrm{p}$ to another particle. For a discussion on the effect of particle distribution regularity, consult \cite{osnes2019}. Furthermore, the particles are not allowed to touch the boundaries of the computational domain. This gives a number of particles ranging between approximately 900 and 1400, depending on the given radius of curvature. For each radius of curvature, five simulations with different particle distributions are conducted to reduce the effect of fluctuations introduced by the particle distribution realizations. The effect of the number of realizations considered is discussed in \cref{sec:EC}. 

Each simulation is initiated as a diverging shock tube. The initial state consists of a high-pressure, high-density region and a region containing air at atmospheric conditions. These are separated by a discontinuity, located $0.156L$ upstream of the particle cloud. The driver section conditions are $p^0=3.6619$ MPa, $\rho^0=12.508$ kg/m$^3$, and the gas is initially at rest. This choice of driver section conditions yields a shock wave with Mach number $M=2.6$ followed by a contact discontinuity without any density jump. The flow state behind the incident shock wave in the planar case is used for normalization purposes and is denoted by the subscript $IS$. Based on these conditions, the particle Reynolds number $Re_p=\rho_{IS} u_{IS}D_p/\mu_{IS}\simeq 5000$. The position of the initial gas state discontinuity is chosen such that the time when the shock wave arrives at the outer shell edge coincides with the arrival time of the head of the rarefaction wave at the inner boundary of the computational domain. A symmetry boundary condition is employed at the inner and axial boundaries and periodic boundary conditions are used in the azimuthal direction. 

The results are analyzed using the volume averaged equations (\cref{eq:Vmass,eq:Vmom}). We define averaging volumes spanning the domain in the axial and azimuthal directions, with a radial extent of $L/60$. The flow quantities are averaged over these bins and over the ensemble of simulations at the same radius of curvature. A timescale based on the initial shock wave velocity and layer thickness 
\begin{equation}
\tau_L = L\left(M\sqrt{\gamma\frac{p^0}{\rho^0}}\right)^{-1},
\label{eq:timescale}
\end{equation}
is used to compare the simulation results.

\section{GRID AND ENSEMBLE CONVERGENCE}
\label{sec:grid-conv}

\subsection{Grid convergence}
\label{sec:grid}
\begin{table}
	\caption{Control volume length scale ($\Delta_\mathrm{CV}$) and total number of control volumes used in the grid study ($N_\mathrm{CV}$).}
	\centering
	\begin{tabular}{c c c}
		$\Delta_\mathrm{CV}\ [\mu$m] & $D_\mathrm{p}/\Delta_\mathrm{CV}$ & $N_\mathrm{CV}$\\
		\hline
		7.5 &$8.4$ &$3.99\times10^6$ \\  
		5    &$12.6$ &$13.5\times10^6$ \\ 
		3.35 &$18.8$ &$41.3\times10^6$ \\
		2.25 &$28.0 $&$14.5\times10^7$ \\
		1.5  &$42.0 $&$49.8\times10^7$\\
		\hline
	\end{tabular}
	\label{tab:gridstudy}
\end{table}

\begin{figure}
	\centering
	\includegraphics[]{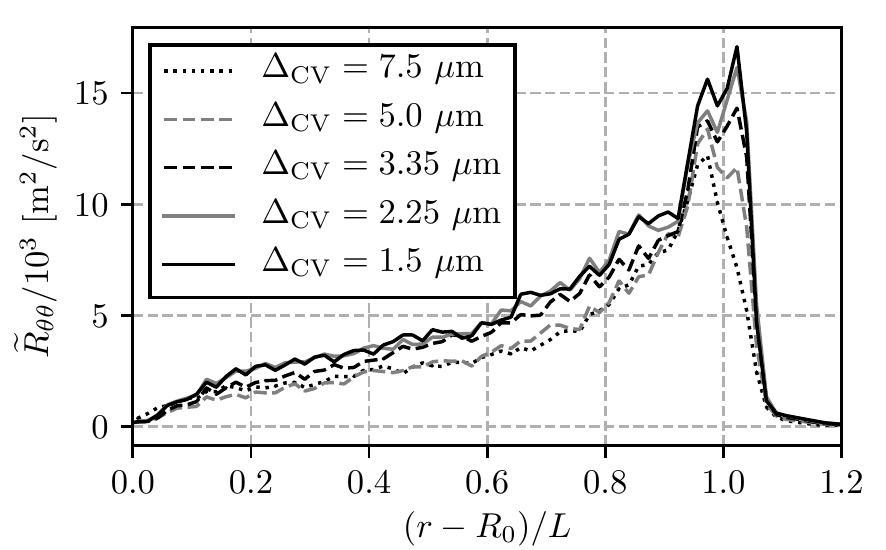}
	\caption{Grid convergence results for $\Rtt$ in the case $R_0=L$ at $t/\tau_L=1.5$. }
	\label{fig:GC}
\end{figure}
To assess the effect of grid sizes, we perform a grid convergence study for $R_0=L$. This study uses a slightly different set-up than the results presented below, with an initial condition corresponding to a $M=2.6$ shock wave located at $r=R_0$ and a corresponding inflow at $r=R_0-0.9L$. The flow pattern for this configuration is similar to the configuration described in \cref{sec:problemsetup} for the initial part of the simulation and we therefore expect the grid dependence to be similar in both cases. Five different grids were utilized, and each new grid had a length scale of roughly $2/3$ of the previous grid length scale. The grid length scales and total number of control volumes are listed in \cref{tab:gridstudy}. These grid length scales are used for regions within a distance $0.5D_\mathrm{p}$ from any particle. Outside of those regions, the grid length scale is doubled, and for $r>R_0+1.5L$ the length scale is doubled once more.

The volume averaged quantities in \cref{eq:Vmass,eq:Vmom} that have the strictest requirement on the computational grid are the Reynolds stresses and the particle forces. By inspection, we find that the azimuthal component of the Reynolds stress is the slowest to converge. \Cref{fig:GC} shows the convergence of $\tilde{R}_{\theta\theta}$ at $t/\tau_L=1.5$. We do not achieve completely converged results within the range of grid length scales used here. Due to the extreme computational cost for the simulations with the finest length scale, we find it necessary to choose $\Delta_\mathrm{CV}=2.25\ \mu \mathrm{m}$ for the simulations within this work. By extrapolating the trend observed in \cref{fig:GC}, we expect the converged Reynolds stresses to be slightly higher than those obtained in our simulations.

\subsection{Ensemble convergence}\label{sec:EC}
In order to minimize the effect of random particle distribution fluctuations, we perform multiple simulations for each $R_0$ and average the results over the simulation ensemble. We determine the number of simulations needed to achieve reasonably converged results based on how $\tilde{R}_{\theta\theta}$ converges with the number of simulations in the case with $R_0=L$. This estimate is based on a grid with $\Delta_\mathrm{CV}=3.35\mu$m, which is coarser than that used for the final simulations. 

\begin{figure}[]
\centering
\begin{subfigure}[t]{.47\textwidth}
	\centering
	\includegraphics[]{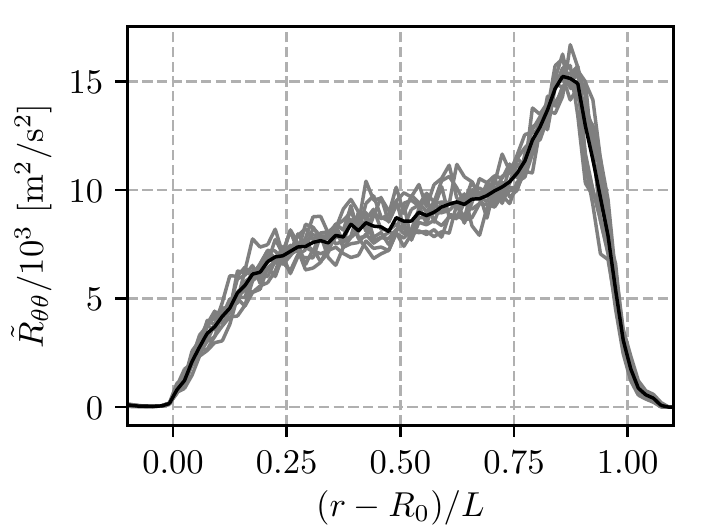}
	\caption{$\tilde{R}_{\theta\theta}$ at $t/\tau_L=1.5$ for the ten different simulations (gray) and their mean value (black).}
	\label{fig:Rtt-realizations}
\end{subfigure}
\hspace{1em}
\begin{subfigure}[t]{.47\textwidth}
	\centering
	\includegraphics[]{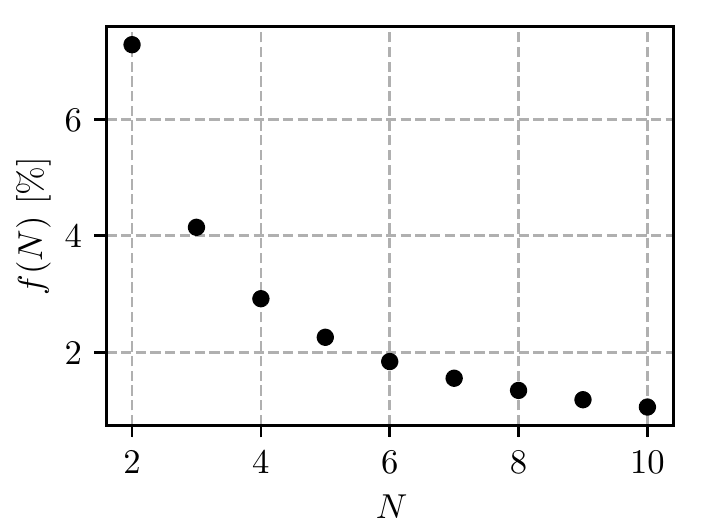}
	\caption{Relative difference between averaging $\tilde{R}_{\theta\theta}$ over $N$ and $N-1$ simulations.}
	\label{fig:ensemble-convergence}
\end{subfigure}
\caption{Ensemble convergence results.}
\end{figure}

\Cref{fig:Rtt-realizations} shows $\Rtt$ at $t/\tau_L=1.5$ for ten different realizations. It is clear that the general shape of all the curves is quite similar and impressions about the trend can be obtained from single simulations. The variation is up to one third of the mean value, and there is a lot to gain from performing an ensemble average.

We quantify the convergence by examining the relative change as we go from $N$ to $N+1$ simulations. This is expressed by the function
\begin{equation}
f(N) = \left[\frac{1}{2.9L}\int_{R_0-0.9L}^{R_0+2L}\left(\frac{\frac{1}{N}\sum_{i=1}^{N}\tilde{R}_{\theta\theta}^i(r)-\frac{1}{N-1}\sum_{i=1}^{N-1}\tilde{R}_{\theta\theta}^i(r)}{\frac{1}{N-1}\sum_{i=1}^{N-1}\tilde{R}_{\theta\theta}^i(r)}\right)^2dr\right]^{1/2}.
\end{equation} 
We compute $f(N)$ for every permutation of the simulation order, and average the results over these combinations. The results are shown in \cref{fig:ensemble-convergence}.
At five realizations, the average relative change in $\tilde{R}_{\theta\theta}$ from adding one more simulation is about $2\%$. At ten realizations, it is $1\%$. Due to the high computational cost and relatively small gain in increasing the number of simulation beyond this point,  we use $N=5$ for all subsequent results presented in this study.  

\section{RESULTS}
\label{sec:results}
\begin{figure}
	\centering
	\includegraphics[]{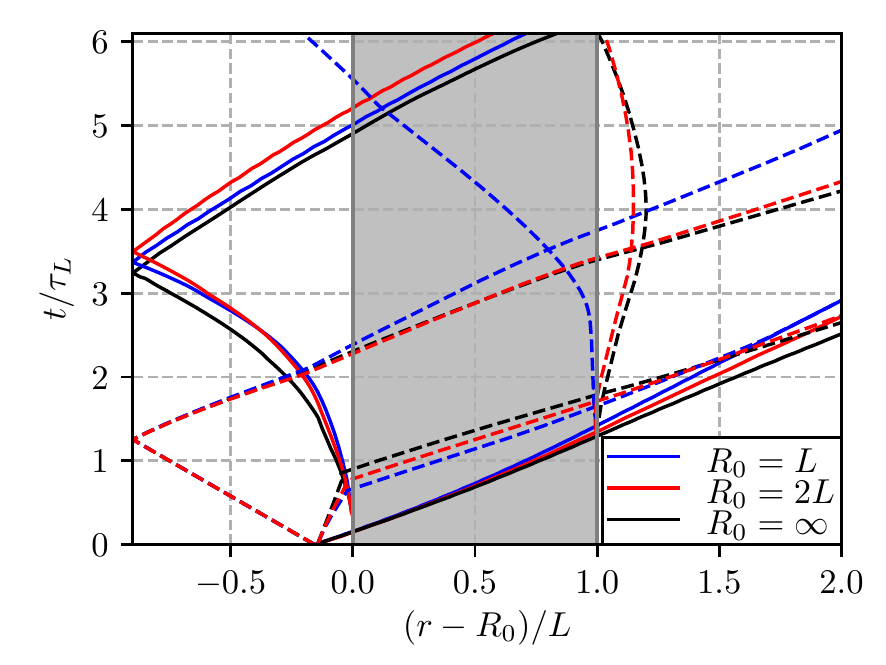}
	\caption{$x-t$ diagram showing the propagation of the primary shock wave (solid), acoustic signals resulting from the head and tail of the initial rarefaction (dashed), reflected shock wave generated at the upstream particle cloud edge (solid) and acoustic signals generated by the shock wave exiting the particle cloud (dashed).}
	\label{fig:xt-diagram}
\end{figure}
Due to similar initial conditions, the simulations for the different radii of curvature display the same basic flow pattern. The discontinuity at the outer boundary of the driver section generates an outward moving shock wave and a rarefaction wave with a head moving inwards and a tail moving outwards. As the shock wave impacts the particle cloud, a reflected shock wave is set up. Subsequently, the tail of the rarefaction also interacts with the upstream boundary of the particle cloud. Thereafter, the head of the rarefaction interacts with the cloud after having reflected off the inner boundary. Finally, the reflected shock wave interacts with the cloud after having reflected off the inner boundary. An $x-t$ diagram showing the position of the above mentioned waves, obtained numerically, is found in \cref{fig:xt-diagram}. This figure also contains the acoustic characteristic $\favg{u}_r-c$ generated as the shock wave exits the particle cloud. The right-ward trajectory of these signals for $R_0=2L$ and $R_0=\infty$ indicates that the flow becomes supersonic immediately at the downstream edge, but we do not observe the same phenomenon for $R_0=L$. 

The interaction of the shock wave with the particle cloud results in a continuous weakening of the shock that slows it down. The geometric expansion also attenuates the shock and slows it down further. If we correct for the weakening of the shock wave due to the geometric expansion, by comparing with simulation results obtained without particles, no significant difference in shock weakening between the three curvature radii is observed.

Based on the wave system described above, we chose $t/\tau_L=0.75$, $1.5$, $3$ and $4.5$ to compare the results for the different cases. At $t/\tau_L=0.75$ the state inside the particle cloud is only affected by the initial shock wave, while at $t/\tau_L=1.5$ both the tail of the rarefaction and the shock wave are involved. At $t/\tau_L=3$, the head of the rarefaction is part way through the cloud. Finally, at $t/\tau_L=4.5$ the flow has developed further but is not yet affected by the reflected shock wave. For the last two times an expansion region is present at the downstream edge of the particle cloud. The strength of this expansion is important for the process of particle dispersion. Its dependence on the radius of curvature will also be discussed below. 

\begin{figure}
	\includegraphics[]{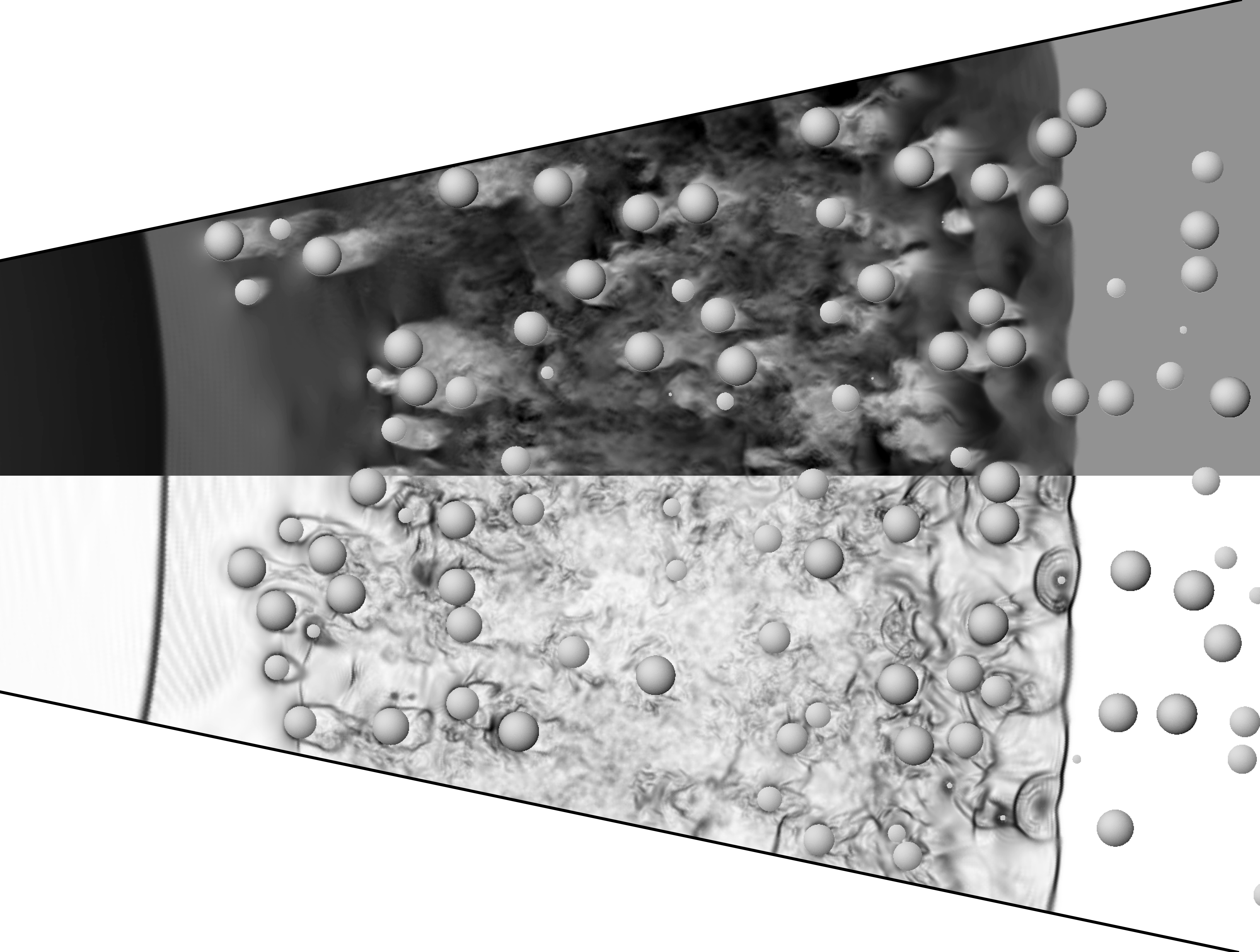}
	\caption{Flow snapshot for $R_0=L$ at $t/\tau_L=1$. The top half displays the radial velocity ($u_r$) and the bottom half displays density gradients using numerical schlieren.}
	\label{fig:flow_viz38}
\end{figure}

\Cref{fig:flow_viz38} contains a visualization of the radial velocity and density gradients within the particle cloud at $t/\tau_L=1$. It illustrates the complexity of the flow field resulting from the interaction. The refracted initial shock wave is visible to the right and directly behind it, the reflected shock waves from the particles. Further upstream, we see the development of particle wakes and shear layers. Upstream of the particle cloud, the reflected shock wave is clearly visible. 

\subsection{Mean flow}
\label{sec:mean}
\begin{figure}[h!]
\includegraphics[]{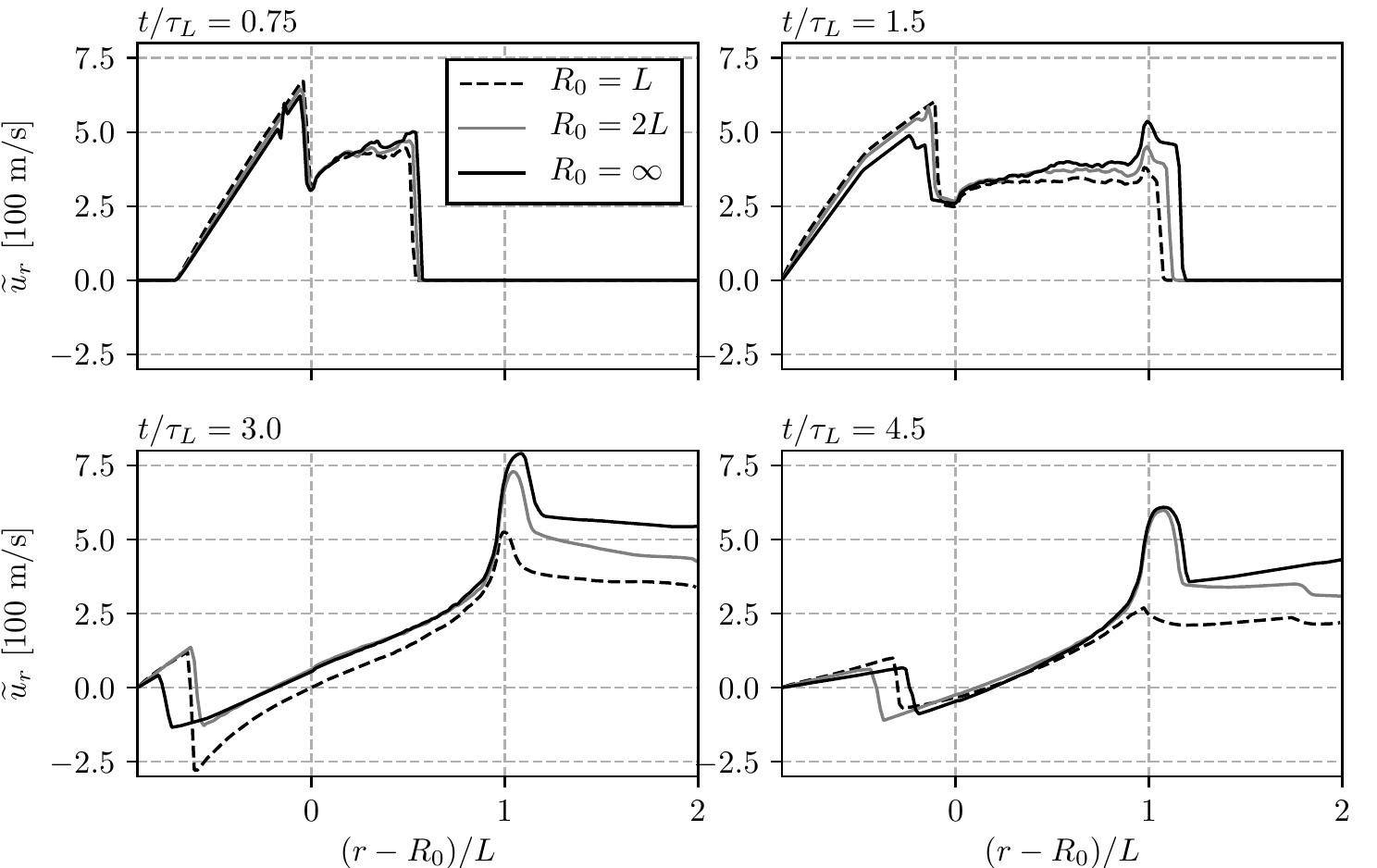}
\caption{Mean radial velocity as a function of $r$ at $t/\tau_L=0.75,\ 1.5,\ 3.0$ and $4.5$. }
\label{fig:ur}
\end{figure}

In this section we examine the mean flow fields for the three curvature radii at the four times $t/\tau_L=0.75$, $1.5$, $3.0$ and $4.5$. \Cref{fig:ur} shows the mean radial velocity. The flow velocity increases with distance within the particle cloud. At the two earliest times, there is a steep gradient just downstream of the inner cloud edge. This is expected since the flow is locally subsonic when it enters the cloud due to the reflected shock. Therefore, the area contraction at the particle cloud edge causes a flow acceleration. It can be seen that the region upstream of the reflected shock has higher radial velocities for smaller $R_0$ due to the geometric expansion. A small numerical artifact can be seen at about $r-R_0\approx-0.2$, which is a result of the discontinuous initial condition on the Voronoi-grid. This effect is present also for smaller $R_0$, but is dampened much faster in those simulations. 

At $t/\tau_L=1.5$, the shock wave has exited the particle cloud, and it can be seen that the flow accelerates towards the downstream cloud edge. The head of the rarefaction wave has reflected off the inner domain boundary, and can be seen as a kink in the velocity profiles around $(r-R_0)/L\approx-0.5$. At later times, the flow slows down due to the reflected rarefaction wave. Indeed, the gas flows inwards in parts of the cloud at $t/\tau_L=3.0$. We also observe that the expansion region at the downstream edge is stronger for larger $R_0$ and persists for a long time. At $t/\tau_L=4.5$, the reflected shock wave is moving outwards, having reflected off the inner boundary, and is visible at $(r-R_0)/L\approx-0.25$.

\begin{figure}
		\includegraphics[]{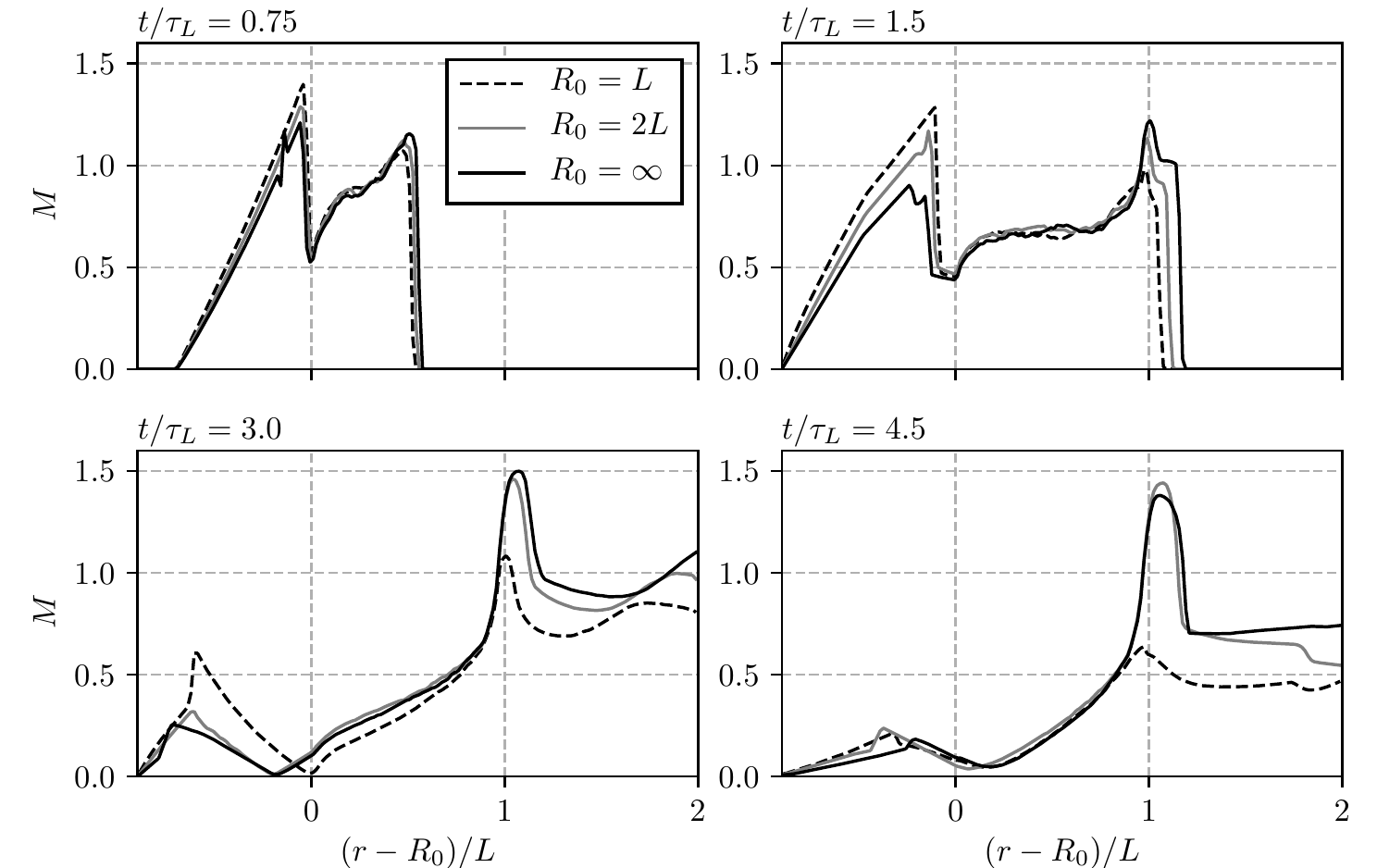}
	\caption{Local Mach number as a function of $r$ at $t/\tau_L=0.75,\ 1.5,\ 3.0$ and $4.5$.}
	\label{fig:M}
\end{figure}

The local flow Mach numbers are shown in \cref{fig:M}. At $t/\tau_L=0.75$ and $t/\tau_L=1.5$,  the region inside the particle cloud has locally higher Mach numbers for smaller curvature radii. In contrast, the local Mach number is larger for larger curvature radii at the downstream cloud edge. For all cases, there is a transition to supersonic flow which occurs just before the edge of the of the particle layer. For $R_0=L$, the flow only becomes supersonic for a very limited time, with a Mach number just above one at $t/\tau_L=3.0$, while for larger $R_0$ local Mach numbers up to 1.5 are observed for extended periods of time. For the latter  cases, we observe that the expansion region is terminated by a quasi-steady shock located at $(r-R_0)/L\approx 1.25$ at the two latest times. 

\begin{figure}
		\includegraphics[]{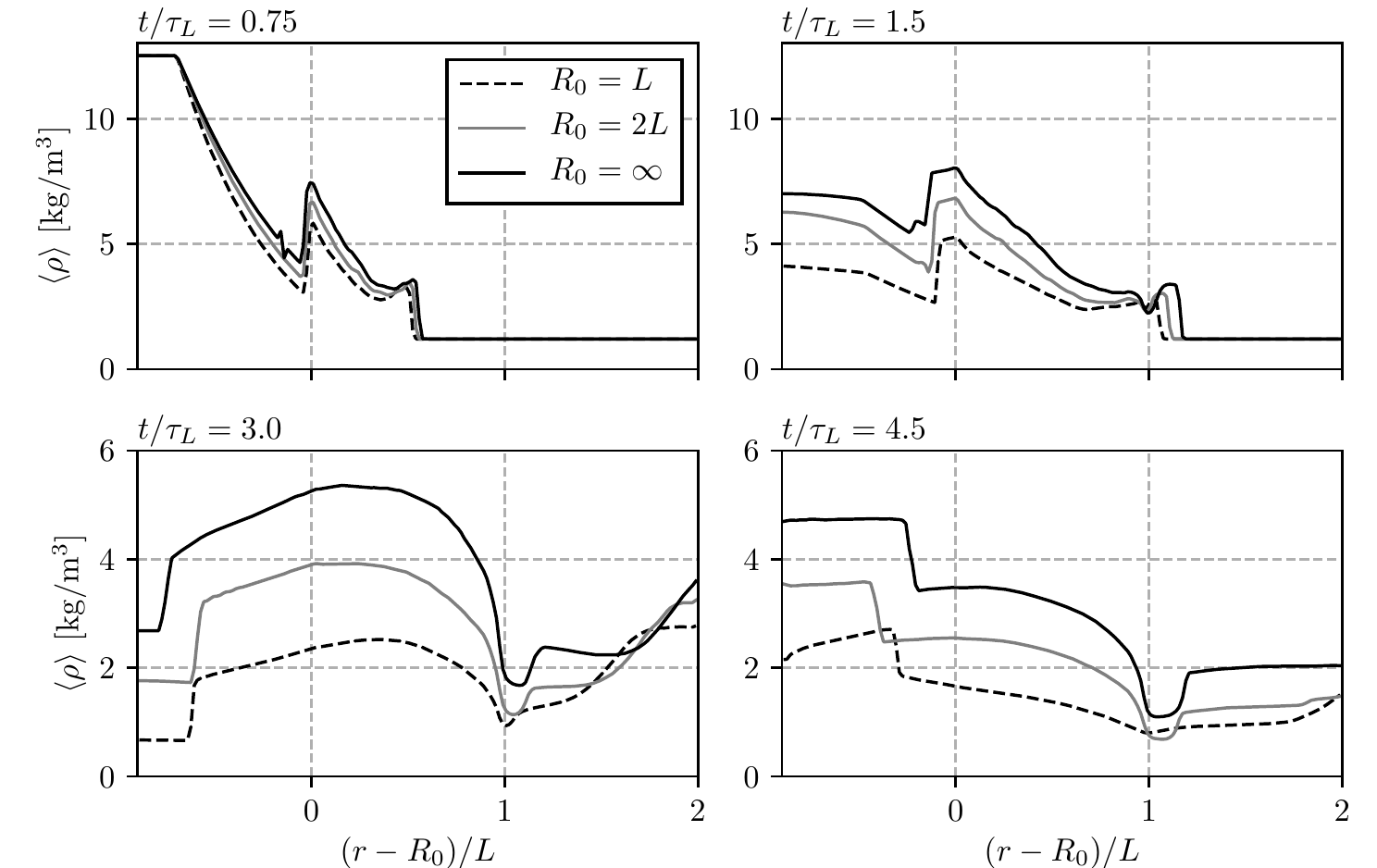}
		\caption{Mean density as a function of $r$ at $t/\tau_L=0.75,\ 1.5,\ 3.0$ and $4.5$. Note the difference in scaling of the vertical axis for the top and bottom.}
		\label{fig:rho}
\end{figure}

The density profiles are shown in \cref{fig:rho}. When the gas expands outwards the mass is distributed over relatively larger volumes for smaller $R_0$. This leads to lower mass densities. It can be seen that the relative jump in density over the reflected shock is higher for lower $R_0$, which means that the reflected shock wave is stronger for smaller curvature radii. This is in accordance with the path of the reflected shocks in \cref{fig:xt-diagram}, where it can clearly be seen that the reflected shock for $R_0=L$ accelerates strongly until it reflects off the inner domain boundary. 

\begin{figure}
		\includegraphics[]{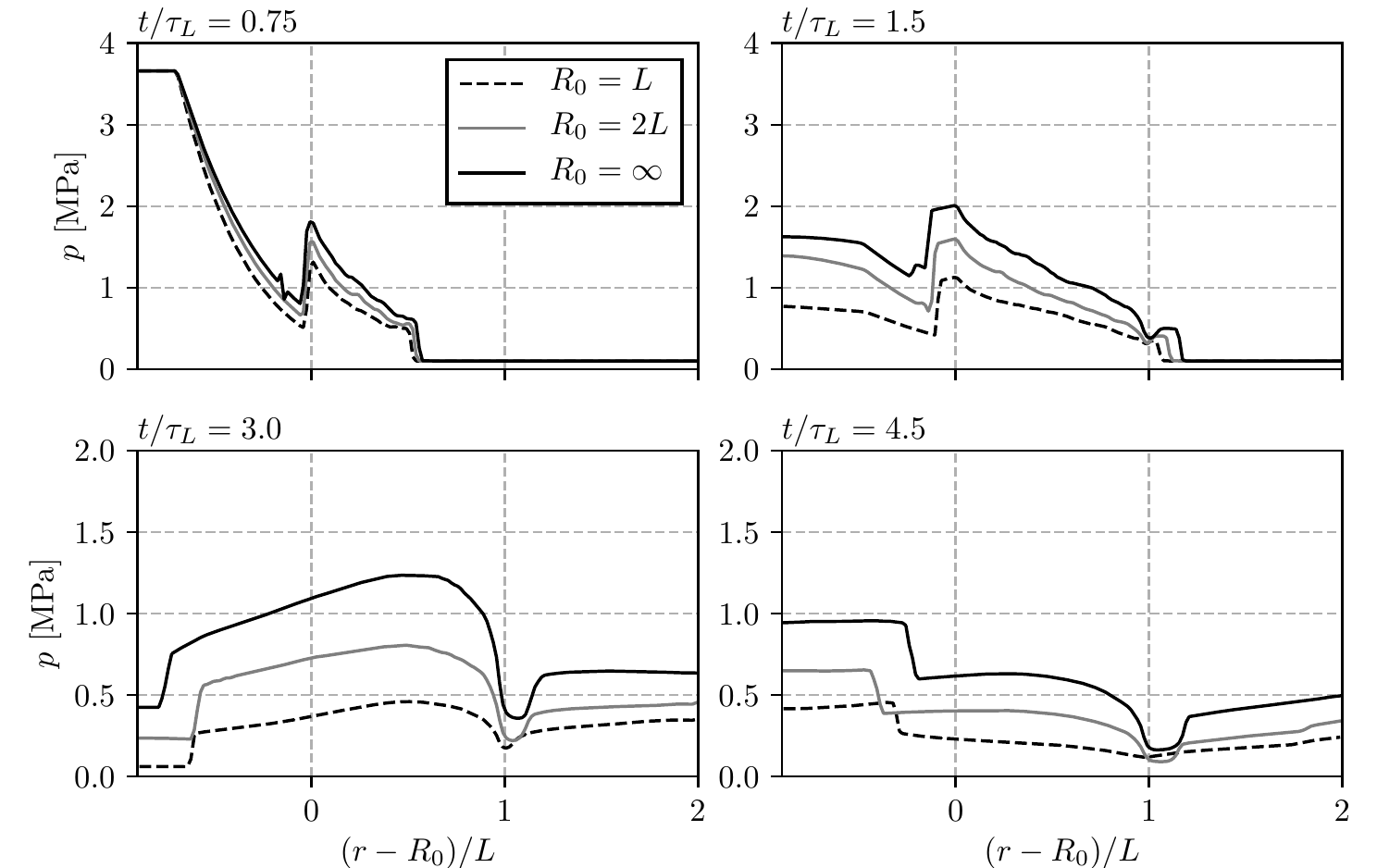}
	\caption{Mean pressure as a function of $r$ at $t/\tau_L=0.75,\ 1.5,\ 3.0$ and $4.5$. Note the difference in scaling of the vertical axis for the top and bottom.}
	\label{fig:p}
\end{figure}

The pressure profiles, found in \cref{fig:p}, display much of the same properties as the density profiles. There is an almost constant gradient through the particle layer after the shock wave has passed, and this state lasts until the rarefaction wave has reflected off the inner boundary and begins to decelerate the flow within the particle cloud. The pressure drops sharply in the expansion at the downstream cloud edge. This is most prominent for larger curvature radii. 

\subsection{Particle forces}
\label{sec:particledrag}

\begin{figure}	
		\includegraphics[]{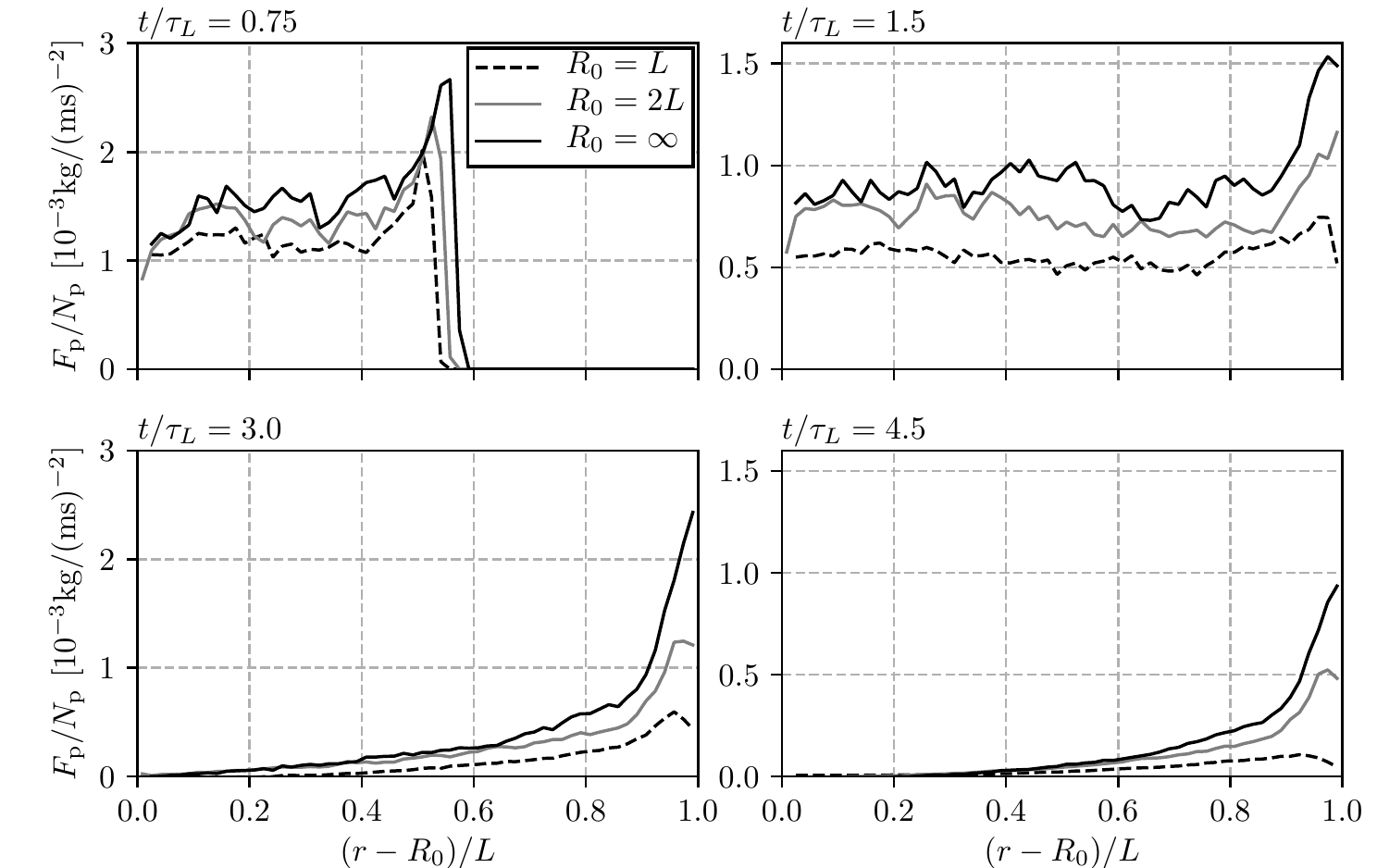}
		\caption{Average force on the particles as a function of $r$ at $t/\tau_L=0.75,\ 1.5,\ 3.0$ and $4.5$. Note the difference in scaling of the vertical axis for the left and right columns.}
		\label{fig:ftr}
\end{figure}

Average particle forces are shown in \cref{fig:ftr}. Smaller $R_0$ lead to lower particle forces, primarily as a result of the lower mean kinetic energy of the flow. At $t/\tau_L=0.75$, it is clear that the largest forces on the particles are imposed during their interaction with the shock wave. Subsequently, the forces tend to a roughly constant value through the particle cloud, except for the drastic increase at the downstream cloud edge. This state lasts until the passage of the rarefaction wave. Interestingly, for $R_0=L$ and $R_0=2L$ there is a slight dip in the forces on the particles towards the end of the particle cloud at the later time intervals. This is in accordance with the distribution of the pressure gradient, which is steepest at the location of the peak particle forces. At the end of the particle layer, the pressure gradient is gentler and therefore contributes less to the forces on the particles. 

\begin{figure}
		\includegraphics[]{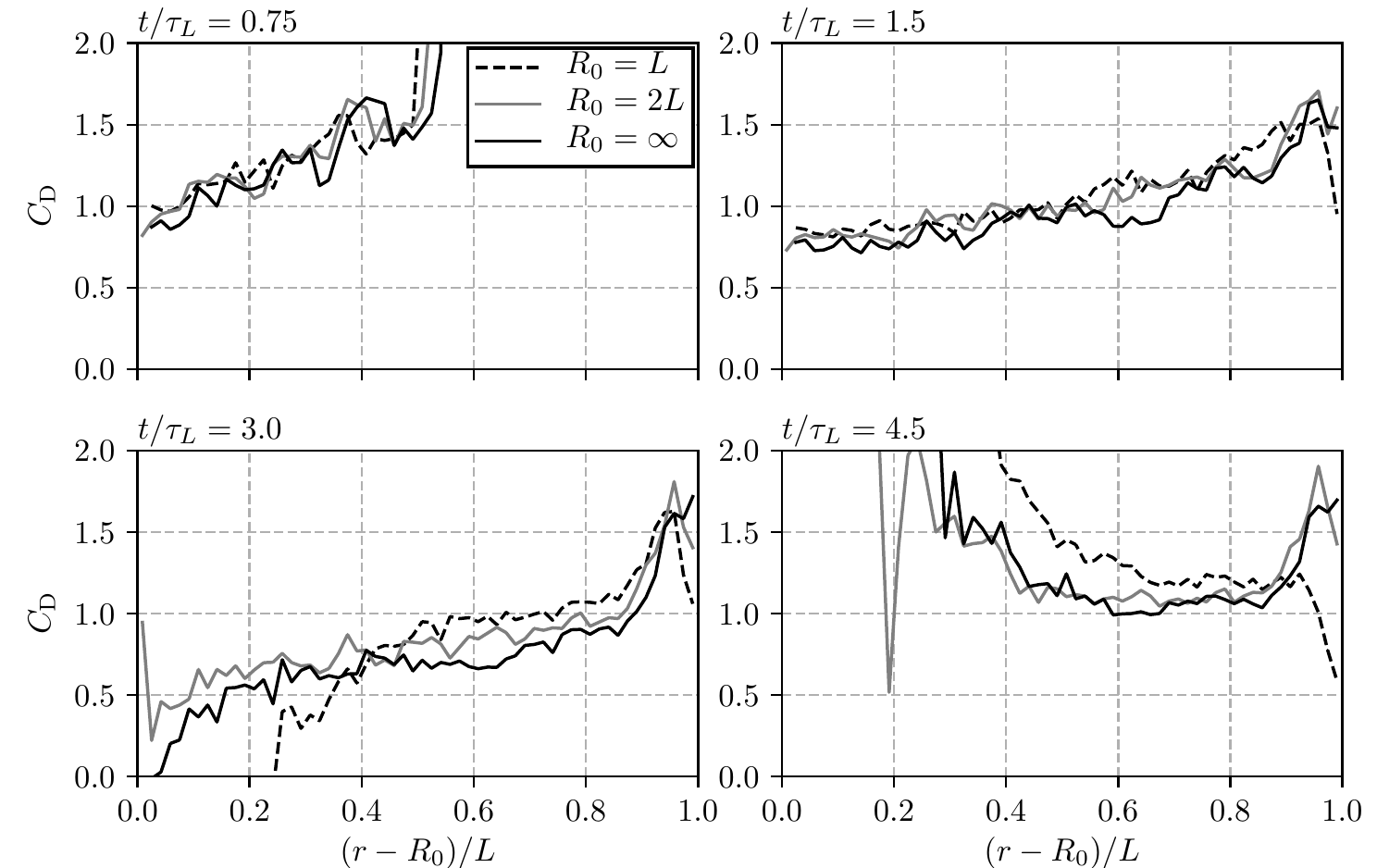}
	\caption{Mean drag coefficient as a function of $r$ at $t/\tau_L=0.75,\ 1.5,\ 3.0$ and $4.5$.}
	\label{fig:Cd}
\end{figure}

The drag coefficients, shown in \cref{fig:Cd}, also increase at the downstream particle cloud edge. The particle drag coefficient is defined as
\begin{equation}
C_D = \frac{\int_{S_i} (-p\delta_{rk}+\sigma_{rk})n_kdS_i}{0.5\phavg{\rho}\favg{u}_r^2A_\mathrm{p}},
\label{eq:Cd}
\end{equation}
where $\delta_{ij}$ is the Kronecker delta, $A_\mathrm{p}$ is the projected area of the particle in the direction of the flow, and $S_i$ denotes the surface of the particle. The increase in drag coefficient signifies that the increase in particle forces is not merely an effect of an increased kinetic energy of the flow. Instead, the increase is a result of the increased Mach number, cf. \cref{fig:M}. This is consistent with findings in studies of single-particle drag as a function of Mach number \cite{bailey1971,nagata2016}.

In contrast to the particle forces, the drag coefficient does not vary much with curvature radius. The minor variation is consistent with standard drag correlations, which predict decreasing $C_\mathrm{D}$ with increasing particle Reynolds numbers.  

Compared to the isolated particle case, the average drag coefficients obtained here are significantly higher. This has also been observed in previous studies \cite{mehta2018,osnes2019}. These studies have shown that there is a wide distribution of drag coefficients, centered higher than isolated particle drag correlations predicts, for particle clouds consisting of randomly distributed particles. In addition to the Mach number effect, a likely contributing factor to this result is the flow blockage effects of nearby particles. Blockage effects have experimentally been shown to significantly increase drag for single particles in ducts \cite{achenbach1974,krishnan2010}.

\subsection{Velocity fluctuations}
\label{sec:velocity-fluctuations}
\begin{figure}
		\includegraphics[]{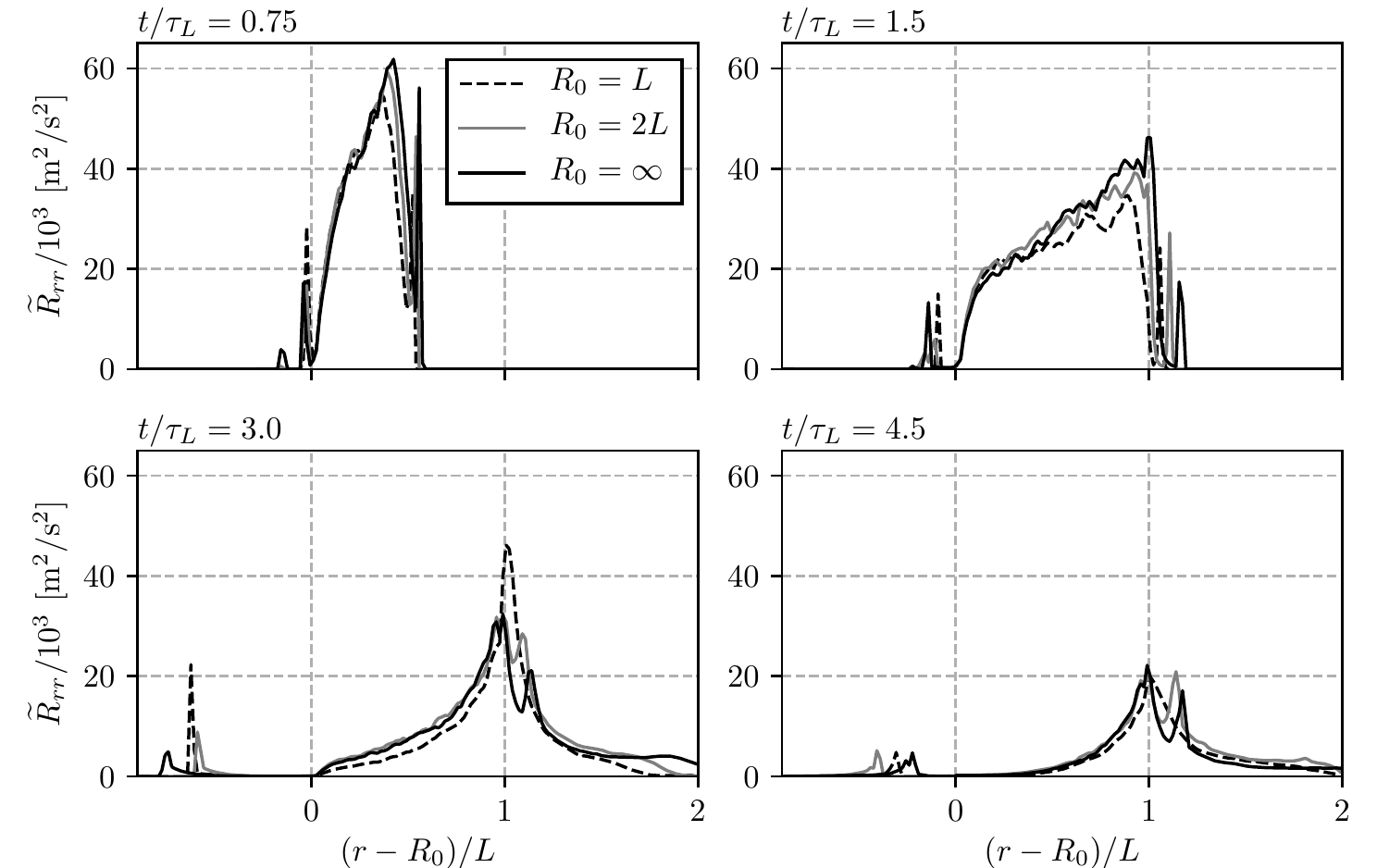}
	\caption{Radial Reynolds stress as a function of $r$ at $t/\tau_L=0.75,\ 1.5,\ 3.0$ and $4.5$.}
	\label{fig:Rrr}
\end{figure}

\Cref{fig:Rrr} shows the radial component of the Reynolds stress. It can be seen that $\Rrr$ increases rapidly immediately behind the shock wave. When the shock wave is inside the particle layer, its peak value occurs a few particle diameters behind the shock wave. After the shock exits the layer, the peak value is at the downstream layer edge. The variation with curvature appears to only manifest in the magnitude of $\Rrr$ at $t/\tau_L=0.75$ and $t/\tau_L=1.5$. At later times this is no longer the case. Two peaks can be seen for $R_0=L$ and $R_0=2L$ at the two latest times. The second peak is the result of the standing shock wave at the end of the expansion. For $R_0=L$, the two peaks merge because this shock wave is closer than one bin-length to the particle layer. The radial component of the Reynolds stress drops sharply over the downstream particle cloud edge. This is expected, because separation in the particle wakes is the primary source of $\Rrr$ in this problem. It does, however, not vanish completely as flow fluctuations are advected downstream from the particle cloud. Interestingly, apart from the previously mentioned shocks, the distribution seems to become more symmetric around the edge at late times.

We note that the magnitude of the streamwise Reynolds stress is significant. It corresponds to root-mean square velocity fluctuations of up to 50\% of the local mean flow velocity. The relation between the mean flow velocity and radial velocity fluctuations will be further discussed below.  

\begin{figure}
		\includegraphics[]{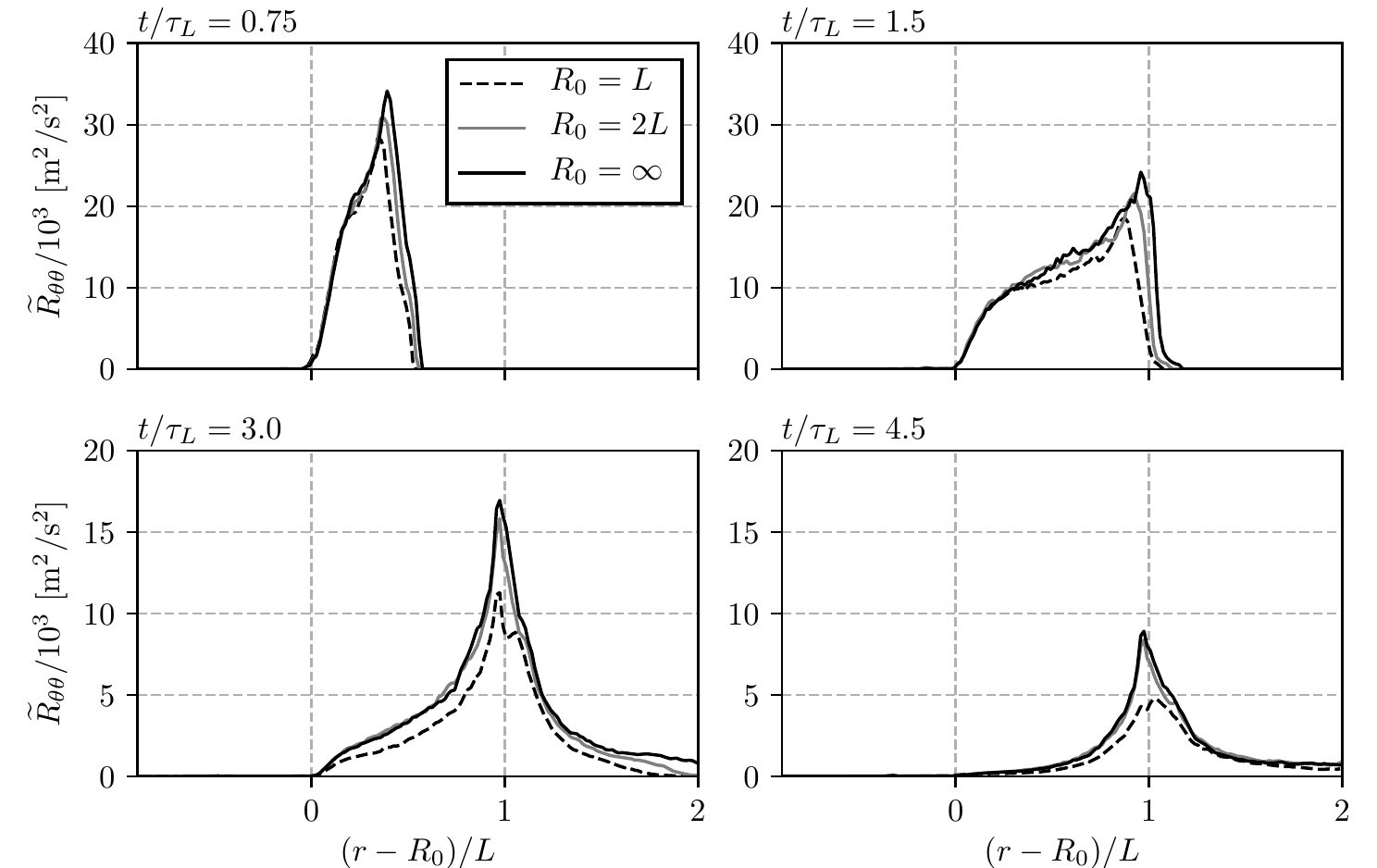}
	\caption{Azimuthal Reynolds stress as a function of $r$ at $t/\tau_L=0.75,\ 1.5,\ 3.0$ and $4.5$. Note the difference in scaling of the vertical axis for the top and bottom.}
	\label{fig:Rtt}
\end{figure}

The azimuthal component of the Reynolds stress, seen in \cref{fig:Rtt}, behaves in much the same way as $\Rrr$. At $t/\tau_L=0.75$ and $t/\tau_L=1.5$, it increases slower with downstream distance than the radial component. Its magnitude is about $50\%$ of $\Rrr$. Surprisingly, the azimuthal and axial (not shown) components of the Reynolds stress are almost indistinguishable. This adds credibility to the claim that the majority of the fluctuations are pseudo-turbulent in nature and that the flow at the particle scale is largely unaffected by the expansion. This result is encouraging from a modeling perspective as it means that results obtained for planar configurations are likely to hold for curvature radii within the range considered here.      

\begin{figure}
		\includegraphics[]{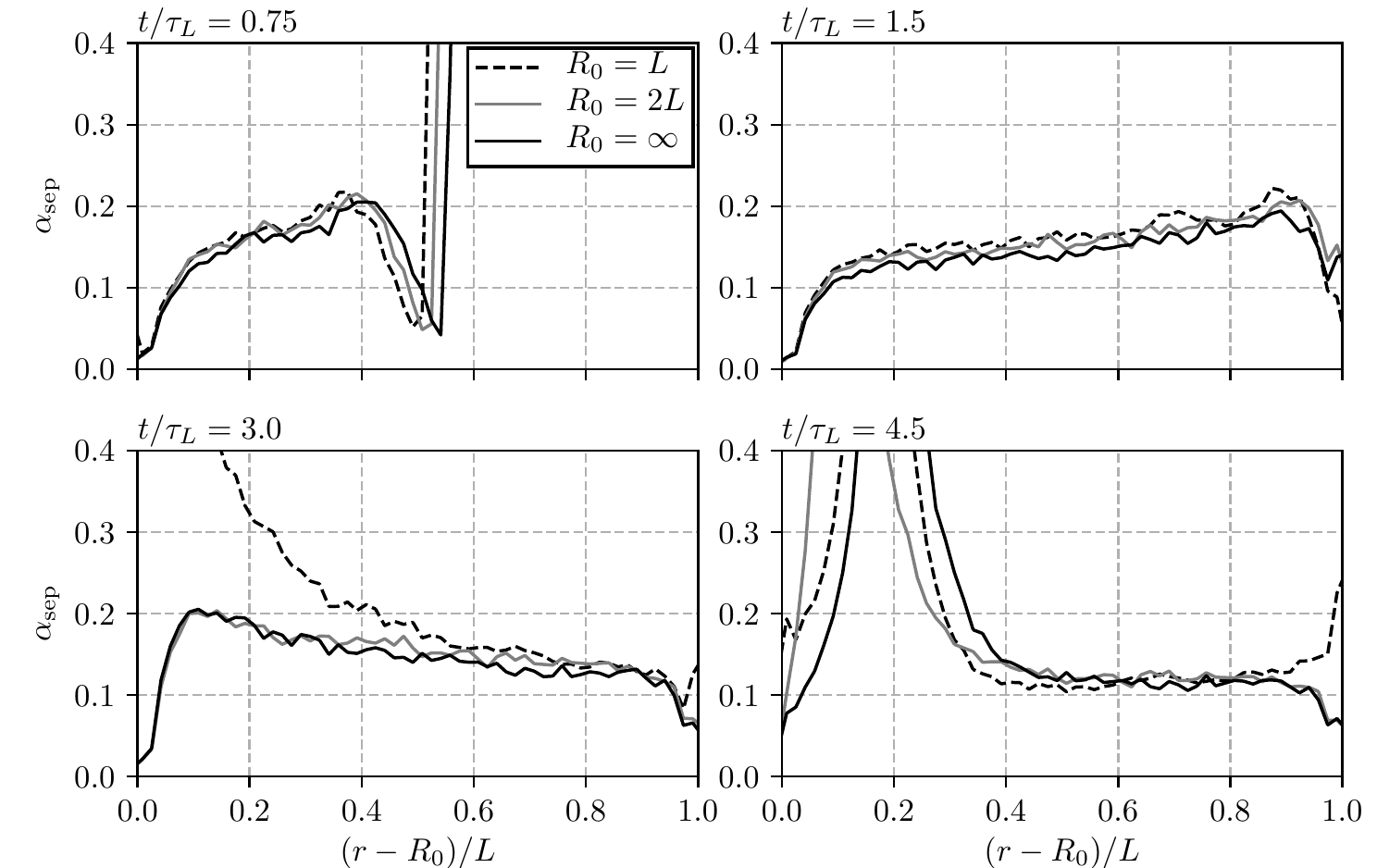}
			\caption{Separation volume as a function of $r$ at $t/\tau_L=0.75,\ 1.5,\ 3.0$ and $4.5$.}
		\label{fig:alphasep}
\end{figure}

It has been observed that the correlation of streamwise velocity fluctuations (when the fluctuations are defined as deviations from volume averages) is approximately proportional to the square of the volume averaged velocity after the strong shock-induced transient has decayed \cite{osnes2019}. It is interesting to investigate whether this is also the case in the current configurations, which differs from the previous study both in domain geometry and initial conditions. The proportionality factor was defined as
\begin{equation}
\alpha_\mathrm{sep}=\alpha\left(1+\widetilde{u}^2/\widetilde{R}_{rr}\right)^{-1},
\label{eq:alphasep}
\end{equation}
and this factor is plotted in \cref{fig:alphasep}. It can be seen that this factor varies only slightly with $R_0$ at the two earliest times. The basis of this model is that the main flow effect that contributes to $\Rrr$ is the separated flow behind each particle. It is clear that this is not a reasonable assumption when the flow decelerates and eventually changes direction, because the particle wakes no longer have the simple behavior required by the model. This effect can be seen at $t \geq 3\tau_L$. It does however appear that as long as the velocity remains fairly high, the model is a decent approximation, cf.  $\alpha_\mathrm{sep}$ at $t/\tau_L=3.0$ for $R_0=2L$ and $R_0=\infty$, and even at $t/\tau_L=4.5$ in the outer half of the particle layer.

The derivation of this model approximates the flow field as two different homogeneous regions. One region is the separated flow behind each particle, where the velocity is approximated as zero. The second region is the flow between particles, where the velocity is assumed to be constant. Improvements to this model can be obtained by analysis of the flow fields around each particle. Such studies could extend the model to include the effects of flow deflection and acceleration around particles, as well as particle volume fraction dependency and particle acceleration. These are topics for future works. 

One of the appealing aspects of this Reynolds stress model is that it is easily applicable to simplified dispersed flow models. It is an algebraic fluctuation model, and thus computationally efficient. Beyond the direct addition of the Reynolds stress, the model also implies corrections to mean flow properties due to the non-negligible volume fraction of the separated flow, as well as the appropriate velocities for computing drag coefficients on the particles. For a more thorough discussion, consult \cite{osnes2019}.

\subsection{Momentum balance}
\label{sec:momentum-balance}

\begin{figure}
		\includegraphics[]{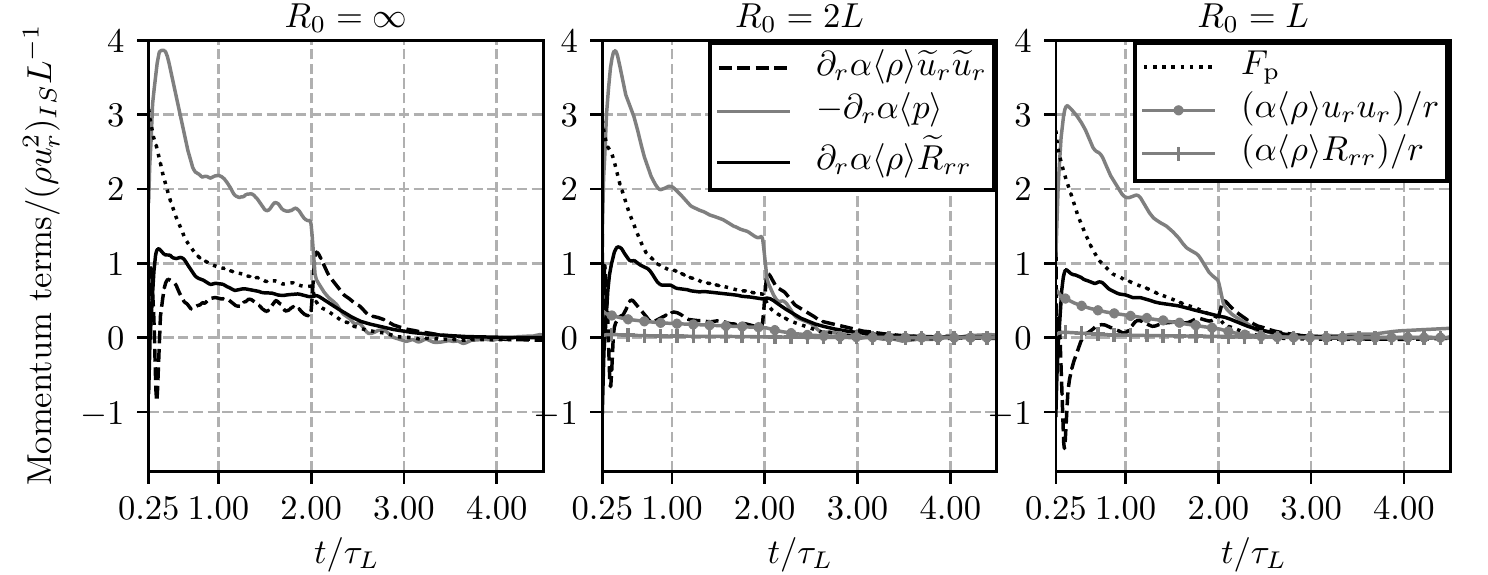}
	\caption{Main terms in the momentum balance close to the upstream edge of the particle cloud ($r=R_0+0.05L$).}
	\label{fig:momentumbalance}
\end{figure}

The time-dependence and radial distribution of flow statistics have been presented above. Here, the relative importance of the terms in the volume averaged momentum balance equation is investigated. \Cref{fig:momentumbalance} shows the most important terms in the momentum balance as a function of time for the three curvature radii at $r=R_0+L/20$. This specific location is chosen for a reason: the momentum balance around the inner cloud boundary has a strong effect on the reflected shock wave. The strength of this shock determines the incoming flow which subsequently interacts with the particle cloud. Therefore, capturing this reflected shock wave correctly is essential for all simplified models for this problem.

It was shown in \cref{fig:Rrr} that around the inner particle cloud edge, the gradient of $\Rrr$ is initially sharp, and it is therefore likely to play an important role in the momentum balance. \Cref{fig:momentumbalance} confirms that this is the case. For all radii of curvature, $\partial_r\alpha\phrho\Rrr$ is the same order of magnitude as the pressure force, $F_p$, acting on the particles. Both of these terms are at least 25\% of the pressure gradient, which is the most dominant. The variation of the Reynolds stress gradient with curvature radius is insignificant. The advection term decreases in importance with decreasing curvature radius, but this effect is partially compensated for by its geometric expansion term. The results indicate that geometric expansion becomes important at small $R_0$, but does not cause any significant changes in flow fluctuations or particle forces. 

It can be seen that for $R_0=\infty$ and $R_0=2L$, there is an abrupt change in the momentum balance around $t/\tau_L=2.0$. The same phenomenon occurs also for $R_0=L$, but the effect is much smaller. This time coincides with the arrival time of the reflected rarefaction, which initiates a strong flow deceleration. This affects mainly the advection term and the pressure gradient. The other terms have a delayed response to the deceleration. 

\begin{figure}
	\centerline{
		\includegraphics[]{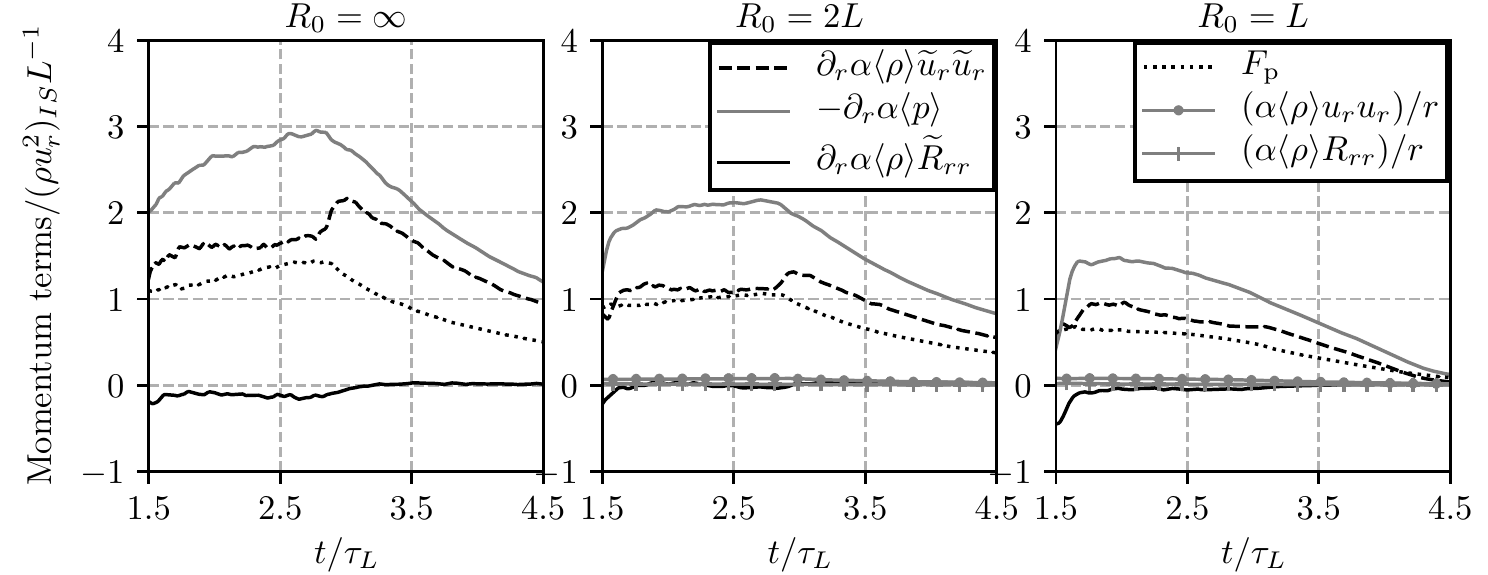}}
	\caption{Main terms in the momentum balance close to the downstream edge of the particle cloud ($r=R_0+0.95L$).}
	\label{fig:momentumbalance_downstream}
\end{figure}

\Cref{fig:momentumbalance_downstream} shows the momentum balance close to the downstream cloud edge. Here, the Reynolds stress is not very important, but we note that this is very close to the peak Reynolds stress, and therefore the gradient is gentler at this point than further downstream. It is clear that the important terms here are the pressure gradient, the advection term and the particle forces. The order of importance of these terms is similar for all $R_0$, and the geometric expansion terms are insignificant.

To complete this discussion, we note that within the interior of the particle cloud, the particle force is the dominant term. The Reynolds stress is non-negligible, but significantly less important than at the upstream cloud edge since its slope is gentler. The pressure gradient and advection terms remain important at all locations.

\section{CONCLUDING REMARKS}
\label{sec:conclusions}

The effect of geometric expansion on shock induced flow through stationary particle clouds has been investigated using viscous particle-resolved simulations in cylindrical domains with differing radii of curvature. 

The main effect of the geometric expansion was found in the mean flow quantities. Density, velocity and pressure were significantly affected, which also translated to differences in the magnitudes of forces experienced by the particles. 
Analysis of the volume averaged momentum balance equations at the upstream particle cloud edge showed that the Reynolds stresses (products of velocity fluctuations) play an important role during the initial part of the flow. During this phase, the Reynolds stress contribution is of the same order as the pressure forces acting on the particles. This means that the Reynolds stresses cannot be ignored in simplified models of shock particle cloud interaction.

The flow fluctuations varied with curvature radius, but mainly due to differences in the mean fields with which they are interacting.
This observation is supported by the fact that no significant difference in the axial and azimuthal components of the Reynolds stresses were observed. This indicates that the geometric expansion rates considered in this work are insufficient to affect the flow at the particle scale. It also supports the hypothesis that the primary contributions to velocity fluctuations are the pseudo turbulent structures at the particle scale.

We also examined whether the Reynolds stress model introduced in \cite{osnes2019} holds for the present configurations. Indeed, a remarkable agreement was obtained for the regions where the mean velocity of the gas was substantial. From a modeling perspective, the results concerning the velocity fluctuations are encouraging because  they indicate that results from planar geometries are likely to hold for diverging flows within the range considered here. Furthermore, the relatively simple correlation between mean flow and Reynolds stress is easy to transfer to simpler dispersed flow models.

Finally, the current results can be used to improve simplified models for shock wave particle cloud interaction. The data from resolved simulations allow comparison of flow fields within the particle cloud. The current data set includes stronger transient effects than e.g. the simulations in \cite{osnes2019}, and can therefore be used as a more challenging verification case for simplified models. 


\begin{thebibliography}{10}
	
	\bibitem{bower1996}
	S.~M. Bower and A.~W. Woods, ``On the dispersal of clasts from volcanic craters
	during small explosive eruptions,'' {\em J. Volcanol. Geotherm. Res.},
	vol.~73, no.~1-2, pp.~19--32, 1996.
	
	\bibitem{suzuki2000}
	K.~Suzuki, H.~Himeki, T.~Watanuki, and T.~Abe, ``Experimental studies on
	characteristics of shock wave propagation through cylinder array,'' tech.
	rep., The Institute of Space and Astronautical Science, 2000.
	
	\bibitem{chaudhuri2013}
	A.~Chaudhuri, A.~Hadjadj, O.~Sadot, and G.~Ben-Dor, ``Numerical study of
	shock-wave mitigation through matrices of solid obstacles,'' {\em Shock
		Waves}, vol.~23, no.~1, pp.~91--101, 2013.
	
	\bibitem{silvia2012}
	D.~W. Silvia, B.~D. Smith, and J.~M. Shull, ``Numerical simulations of
	supernova dust destruction. {I}{I}. {M}etal-enriched ejecta knots,'' {\em
		Astrophys. J.}, vol.~748, no.~1, p.~12, 2012.
	
	\bibitem{zhang2006}
	F.~Zhang, S.~Murray, and K.~Gerrard, ``Aluminum particles--air detonation at
	elevated pressures,'' {\em Shock Waves}, vol.~15, no.~5, pp.~313--324, 2006.
	
	\bibitem{zhang2001}
	F.~Zhang, D.~Frost, P.~Thibault, and S.~Murray, ``Explosive dispersal of solid
	particles,'' {\em Shock Waves}, vol.~10, no.~6, pp.~431--443, 2001.
	
	\bibitem{milne2010}
	A.~Milne, C.~Parrish, and I.~Worland, ``Dynamic fragmentation of blast
	mitigants,'' {\em Shock Waves}, vol.~20, no.~1, pp.~41--51, 2010.
	
	\bibitem{rodriguez2017}
	V.~Rodriguez, R.~Saurel, G.~Jourdan, and L.~Houas, ``Impulsive dispersion of a
	granular layer by a weak blast wave,'' {\em Shock Waves}, vol.~27, no.~2,
	pp.~187--198, 2017.
	
	\bibitem{frost2012}
	D.~L. Frost, Y.~Gr{\'e}goire, O.~Petel, S.~Goroshin, and F.~Zhang, ``Particle
	jet formation during explosive dispersal of solid particles,'' {\em Phys.
		Fluids}, vol.~24, no.~9, p.~091109, 2012.
	
	\bibitem{rodriguez2014}
	V.~Rodriguez, R.~Saurel, G.~Jourdan, and L.~Houas, ``External front
	instabilities induced by a shocked particle ring,'' {\em Phys. Rev. E},
	vol.~90, no.~4, p.~043013, 2014.
	
	\bibitem{frost2018}
	D.~L. Frost, ``Heterogeneous/particle-laden blast waves,'' {\em Shock Waves},
	vol.~28, pp.~439--449, May 2018.
	
	\bibitem{theofanous2017-2}
	T.~G. Theofanous and C.-H. Chang, ``The dynamics of dense particle clouds
	subjected to shock waves. {P}art 2. {M}odeling/numerical issues and the way
	forward,'' {\em Int. J. Multiph. Flow}, vol.~89, pp.~177--206, 2017.
	
	\bibitem{regele2014}
	J.~D. Regele, J.~Rabinovitch, T.~Colonius, and G.~Blanquart, ``Unsteady effects
	in dense, high speed, particle laden flows,'' {\em Int. J. Multiph. Flow},
	vol.~61, pp.~1--13, 2014.
	
	\bibitem{hosseinzadeh2018}
	Z.~Hosseinzadeh-Nik, S.~Subramaniam, and J.~D. Regele, ``Investigation and
	quantification of flow unsteadiness in shock-particle cloud interaction,''
	{\em Int. J. Multiph. Flow}, vol.~101, pp.~186--201, 2018.
	
	\bibitem{sen2018}
	O.~Sen, N.~J. Gaul, S.~Davis, K.~K. Choi, G.~Jacobs, and H.~S. Udaykumar,
	``Role of pseudo-turbulent stresses in shocked particle clouds and
	construction of surrogate models for closure,'' {\em Shock Waves}, vol.~28,
	pp.~579--597, May 2018.
	
	\bibitem{mehta2018}
	Y.~Mehta, C.~Neal, K.~Salari, T.~L. Jackson, S.~Balachandar, and S.~Thakur,
	``Propagation of a strong shock over a random bed of spherical particles,''
	{\em J. Fluid Mech.}, vol.~839, pp.~157--197, 2018.
	
	\bibitem{theofanous2018}
	T.~G. Theofanous, V.~Mitkin, and C.-H. Chang, ``Shock dispersal of dilute
	particle clouds,'' {\em J. Fluid Mech.}, vol.~841, pp.~732--745, 2018.
	
	\bibitem{vartdal2018}
	M.~Vartdal and A.~N. Osnes, ``Using particle-resolved {L}{E}{S} to improve
	{E}ulerian-{L}agrangian modeling of shock wave particle cloud interaction,''
	in {\em Proceedings of the Summer Program 2018}, pp.~25--34, Center for
	Turbulence Research, Stanford University, 2018.
	
	\bibitem{osnes2019}
	A.~N. Osnes, M.~Vartdal, M.~G. Omang, and B.~A.~P. Reif, ``Computational
	analysis of shock-induced flow through stationary particle clouds,'' {\em
		Int. J. Multiph. Flow}, vol.~114, pp.~268 -- 286, 2019.
	
	\bibitem{schwarzkopf2015}
	J.~D. Schwarzkopf and J.~A. Horwitz, ``{B}{H}{R} equations re-derived with
	immiscible particle effects,'' tech. rep., Los Alamos National Laboratory,
	2015.
	
	\bibitem{shallcross2018}
	G.~S. Shallcross and J.~Capecelatro, ``A parametric study of particle-laden
	shock tubes using an {E}ulerian-{L}agrangian framework,'' in {\em 2018 AIAA
		Aerospace Sciences Meeting}, p.~2080, 2018.
	
	\bibitem{mcgrath2016}
	T.~P. McGrath, J.~G.~S. Clair, and S.~Balachandar, ``A compressible two-phase
	model for dispersed particle flows with application from dense to dilute
	regimes,'' {\em J. Appl. Phys.}, vol.~119, p.~174903, 2016.
	
	\bibitem{saurel2017}
	R.~Saurel, A.~Chinnayya, and Q.~Carmouze, ``Modelling compressible dense and
	dilute two-phase flows,'' {\em Phys. Fluids}, vol.~29, no.~6, p.~063301,
	2017.
	
	\bibitem{bres2018}
	G.~A. Bres, S.~T. Bose, M.~Emory, F.~E. Ham, O.~T. Schmidt, G.~Rigas, and
	T.~Colonius, ``{L}arge-{E}ddy {S}imulations of co-annular turbulent jet using
	a {V}oronoi-based mesh generation framework,'' in {\em 2018 AIAA/CEAS
		Aeroacoustics Conference}, p.~3302, 2018.
	
	\bibitem{tadmor2003}
	E.~Tadmor, ``Entropy stability theory for difference approximations of
	nonlinear conservation laws and related time-dependent problems,'' {\em Acta.
		Numer.}, vol.~12, pp.~451--512, 2003.
	
	\bibitem{chandrashekar2013}
	P.~Chandrashekar, ``Kinetic energy preserving and entropy stable finite volume
	schemes for compressible {E}uler and {N}avier-stokes equations,'' {\em
		Commun. Comput. Phys.}, vol.~14, no.~5, pp.~1252--1286, 2013.
	
	\bibitem{bailey1971}
	A.~Bailey and J.~Hiatt, ``Free-flight measurements of sphere drag at subsonic,
	transonic, supersonic, and hypersonic speeds for continuum, transition, and
	near-free-molecular flow conditions,'' tech. rep., Arnold Engineering
	Development Center, 1971.
	
	\bibitem{nagata2016}
	T.~Nagata, T.~Nonomura, S.~Takahashi, Y.~Mizuno, and K.~Fukuda, ``Investigation
	on subsonic to supersonic flow around a sphere at low {R}eynolds number of
	between 50 and 300 by direct numerical simulation,'' {\em Phys. Fluids},
	vol.~28, no.~5, p.~056101, 2016.
	
	\bibitem{achenbach1974}
	E.~Achenbach, ``The effects of surface roughness and tunnel blockage on the
	flow past spheres,'' {\em J. Fluid Mech.}, vol.~65, no.~1, pp.~113--125,
	1974.
	
	\bibitem{krishnan2010}
	S.~Krishnan and A.~Kaman, ``Effect of blockage ratio on drag and heat transfer
	from a centrally located sphere in pipe flow,'' {\em Eng. Appl. Comp.
		Fluid.}, vol.~4, no.~3, pp.~396--414, 2010.
	
\end{thebibliography}

\end{document}